\documentclass[aps,prl,preprint,showpacs,byrevtex]{revtex4}
\usepackage{epsfig}

\begin{document}

\title{Kinematics of stock prices}

\bigskip
\bigskip
\author{ M. Serva$^{(1,2)}$, U. L. Fulco$^{(1)}$, M. L. Lyra$^{(1)}$,
and G. M. Viswanathan$^{(1)}$}
\affiliation{$^{(1)}$Departamento de F\'{\i}sica, Universidade Federal de
 Alagoas,  Macei\'{o}--AL, 57072-970, Brazil}
\affiliation{$^{(2)}$Dipartimento di Matematica and I.N.F.M.
Universit\`a dell'Aquila,
I-67010 L'Aquila, Italy}
\bigskip

\revised{4 September 2002, subm. Phys. Rev. Lett.}

\begin{abstract}

We investigate the general  problem of how to model the kinematics of
stock prices without considering the dynamical causes of motion.  We
propose a stochastic process with long-range correlated absolute
returns. We find that the model is able to reproduce the
experimentally observed clustering, power law memory, fat tails and
multifractality of real financial time series.  We find that the
distribution of stock returns is approximated by a Gaussian with
log-normally distributed local variance and shows excellent agreement
with the behavior of the NYSE index for a range of time scales.

\pacs{05.40.F,89.65.G,05.40.F}

\end{abstract}

\maketitle



\begin{figure}
\vspace{.2in}
\centerline{\psfig{figure=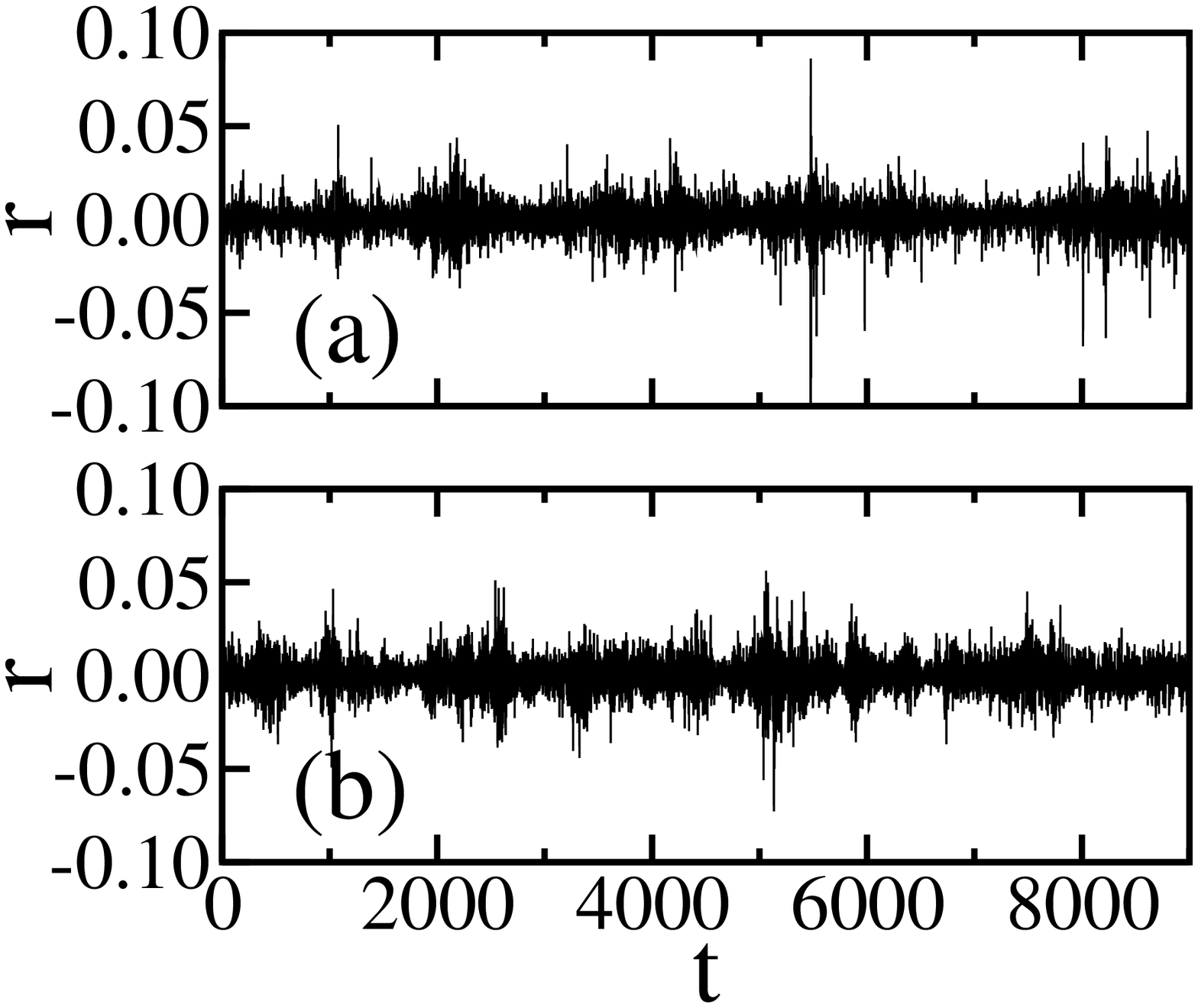,width=5.0truein}}
\bigskip
\caption{Real (a) and artificial data (b) obtained from
the maps in Eqs.~\ref{map1} and \ref{map2} with constants $a=-0.1$, $b=0.1$,
$d=0.98$ and $T=10$. The striking resemblance between the time series
generated by the the kinematic model and the real data is further
verified by quantitative analysis, as explained in the text. }
\label{f1}
\end{figure}

Remarkable progress has been made in quantitatively describing
nonstationary and non-Gaussian phenomena, including those observed in
economic~\cite{econophysicsbook} and social
systems.  The behavior of financial markets has
recently~\cite{econophysicsbook,prl1,plamen1,lrc1,lrc-old0,gopi,lrc2,mantegna2,mantegna}
become a focus of interest to physicists as well as an area of active
research because of its rich and
complex~\cite{bacchelier,ding,crato,bouchaud,levy,baillie,galluccio,vandewall,Liu,volatility-stanley}
dynamics.
The daily returns $r_t$ for a given stock can be defined as
\[
r_t=\log_{10}\frac{S_{t+1}}{S_t}\;\;,
\]
at time $t$ (Fig.~\ref{f1}), where $S_t$ is the price of a an asset or
the value of an index or a currency exchange rate.  In this paper we
address the important yet unsolved problem of how to model the
kinematics of stock prices without handling the problem of the causes
of motion.  Our point of departure is the family of ARCH-GARCH
models. These families of stochastic kinematic models neglect the
causes underlying the price variations and focus only on the equations
of motion governing the fluctuating returns. Specifically, the
interacting buying and selling processes are not considered. ARCH
models, introduced by Engle in 1982~\cite{engle}, are stochastic
Auto-Regressive Conditional Heteroscedasticity models, characterized
by zero mean and nonconstant variances dependent on the past.  Such
models can simultaneously have global stationarity as well as local
nonstationary behavior. The simplest ARCH model of stock returns can
be defined by
\[
\sigma^2_t=\alpha_0 + \alpha_1 r^2_{t-1}\;\;,
\]
with returns defined as
\[
r_t= \sigma_t \omega_t \;\;,
\]
where the random Gaussian variable $\omega_t$ has unitary variance
and zero mean, while $\alpha_0$ and $\alpha_1$ are tunable parameters.
Hence, the returns $r_t$ are Gaussian distributed with zero mean and
local variance $\sigma_t^2.$ Only very recently has the physics
community begun to study such models, although they are common in the
economics literature.  In more general ARCH processes, the local
variance can depend not only on the previous value $r_{t-1}$, but on
any finite number $n$ of previous values $r_{t-1} \ldots r_{t-n}$.
GARCH models are a further generalization in which the local variance
can depend not only on previous values of the returns but also on the
previous values of the locally measured variances.

ARCH-GARCH models have succeeded in capturing important features (such
as volatility clustering) of real financial data.  They are,
therefore, widely used, both in the financial community and in
econophysics to model the kinematics of price
variations. Nevertheless, several characteristic features of the real
data are not well described. The three most important (and difficult
to model) features not well accounted in ARCH-GARCH models are (i) the
time scaling of the probability density distribution of the returns,
(ii) the known long-range power law correlations in the volatility
(the latter being a measure of the local standard deviation of the
returns), and (iii) the multifractality of the returns, i.e.,
volatility power law correlation exponents are non-unique and depend
on the magnitude of the events considered.  As an alternative model,
L\'evy~\cite{mandel} and truncated (see ref.~\cite{tlf}) L\'evy distributions
have been proposed to fit the observed\ fat tailed distributions.
While L\'evy processes can correctly reproduce the time scaling of
distributions, they cannot explain one of the most relevant and
characteristic phenomena: the multifractal long-range correlations in
volatility which has been found to be responsible for clustering of
volatility and the persistence of fat tails for long time
lags~\cite{gopi,gandhi-condmat2002}.

Here we propose a new model that does not suffer from these drawbacks.
In order to test the model, we compare the model with a typical
dataset: the New York Exchange (NYSE) daily composite index price
closes from January 1966 to September 2001 (a total of some 9000 data
points).


 The
model is inspired in the recent finding that the probability density
distribution of the local variance $\sigma_t$ is similar to a
log-normal~\cite{lrc1} . Other key findings
include the long-range correlations found in the absolute value of the
returns and the multifractal behavior of the returns~\cite{lrc2}.  Our
proposal, based on these findings,
is the following map for the
evolution of the volatility:
\begin{equation}
\sigma_{t+1}= e^{[a+b\omega_{t+1}]} ~ (\overline{\sigma_t})^d\;\; , 
\label{map1}
\end{equation}
with returns defined as
\begin{equation}
r_t= \sigma_t ~
\tilde{\omega}_t \;\;,
\label{map2}
\end{equation}
where $\tilde{\omega_t}$ and $\omega_t$ are independently distributed
unitary Gaussian variables with zero mean that are also independent
from each other.  Note that the independence of these two variables
contrasts with ARCH models, for which they coincide.  We discuss why
this is preferable below.  The tunable constants $a$, $b$ and $d$ are
related to the nearly log-normal form of $\sigma_t.$ The value of $d$
must be close but smaller than 1 to guaranty the stationarity of the
returns; $a$ is related to the typical size of the daily volatility,
and $b$ to the typical size of its fluctuations.  The term
$\overline{\sigma_t}$ represents the average with respect to last $T$
days, from $t-T$ to $t$. (We have used below a simple unweighted
average but, in principle, we could have used any type of moving
average, for instance a power law or exponentially weighted moving
average.)

Using this map, we generate a data set of $9000$ returns
(Fig.~\ref{f1}(b)) with the following choice of constants: $a=-0.1$,
$b=0.1$, $d=0.98$ and $T=10$.  We then compare the statistical and
scaling properties of the time series generated with the real data
set.  Note that the above model can reproduce very well the clustering
of volatility, as is evident comparing Fig.~\ref{f1}(a) with
Fig.~\ref{f1}(b).  Nevertheless, we proceed with a quantitative
comparison of both the scaling of the distribution and the
multifractal power law correlations.

\begin{figure}
\vspace{.2in}
\centerline{\psfig{figure=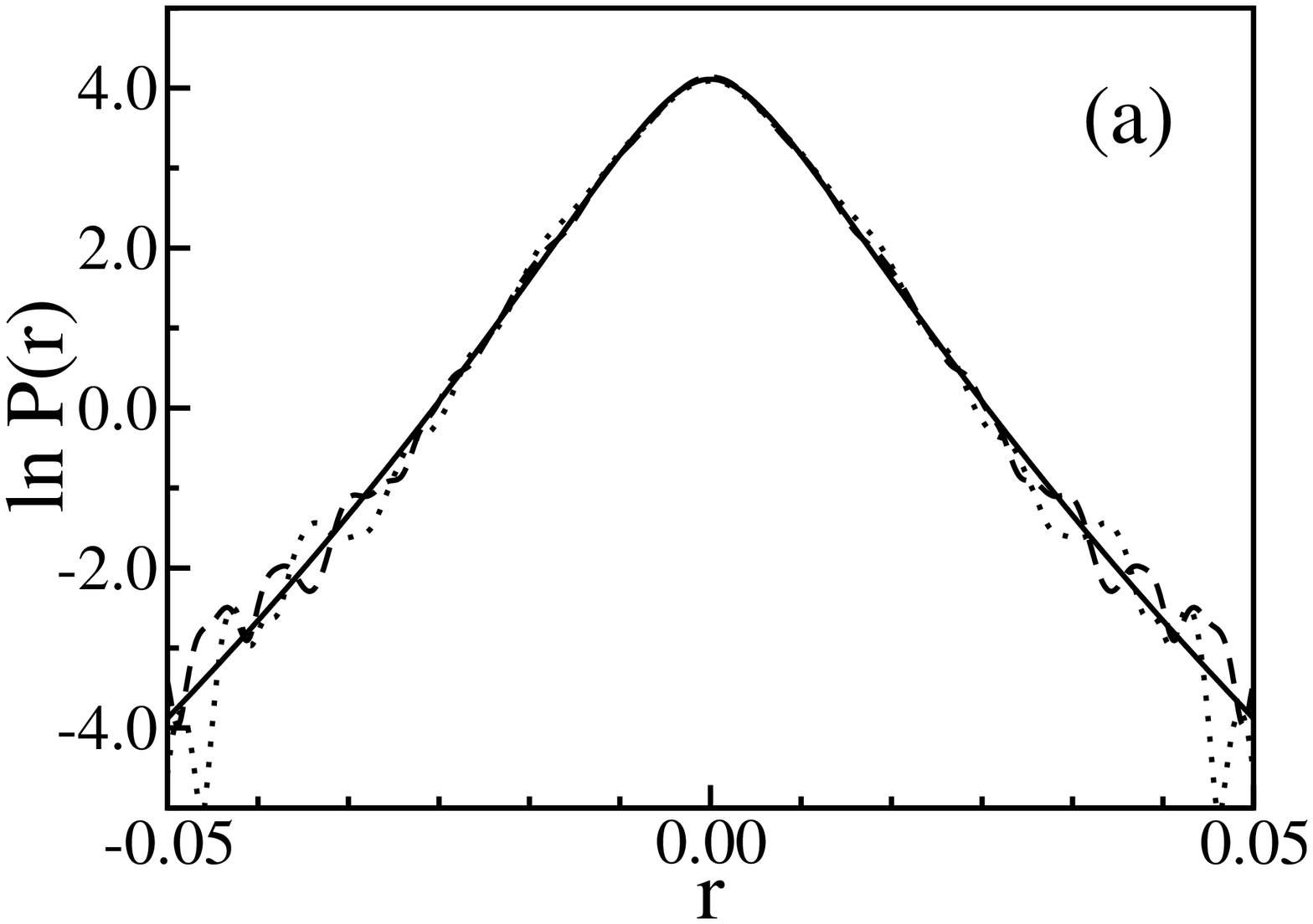,width=4.0truein}}
\centerline{\psfig{figure=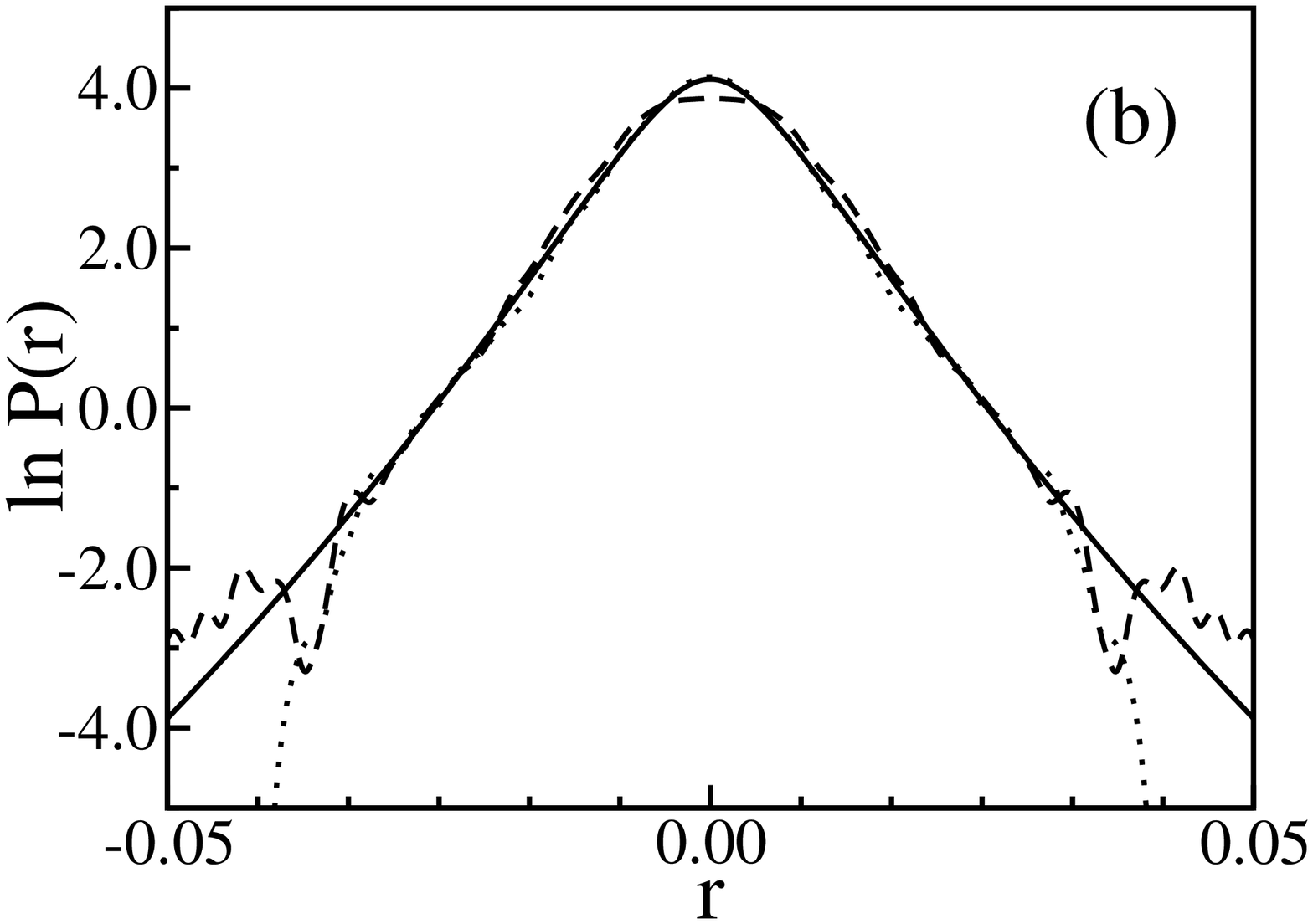,width=4.0truein}}
\bigskip
\caption{Symmetrized probability density distribution $p(r)$ of the
returns $r_{\tau}$measured over lags (a) $\tau=1$ d and (b) $\tau=25 $
d (about one business month), shown for the artificial data (dashed
line), the real data (long dashed) and compared with the
``theoretical'' distribution (solid line)).  For $\tau=1$ d all three
distributions are almost identical while for $\tau=25$ a small
discrepancy appears for real data for small returns. This discrepancy
is due to short range correlations in real data not included in the
model, as explained in the text.}
\label{f2}
\end{figure}

We next study the probability density of returns for the proposed
model.  We estimate the probability density of returns using
\[
P(r)=\frac{1}{N}\sum_{i=1}^{N}\frac{1}{\sqrt{2\pi}\Delta}\exp\{\frac{-(r-r_i)^2}{2\Delta^2}\}
\]
with a smoothing window of $\Delta=0.001$ (Fig~\ref{f2}a). The $r_i$
can be either the real or artificial data and $N$ is their number(about
9000 for both).  The dashed and dotted lines in Fig.~\ref{f2}(a) show
the distributions of daily (one business day) returns, for both real
and artificial data.  Typical of financial times series is their
invariance under rescaling of time.  Therefore, we estimate with the
same smoothing window the distribution for monthly (25 business days)
returns (Fig~\ref{f2}b)
\[
\frac{\sum_{i=1}^{25}r_{i+t}}{(25)^\delta}=\frac{1}{(25)^{\delta}}\log\frac{S_{t+25}}{S_t}\;\; ,
\]
using the measured value $\delta=0.53$ of the scaling exponent both
for real and artificial data.  Note that the theoretical curve (solid
line, see discussion below) is exactly the same in the two figures,
therefore both real and artificial data exhibit almost perfect time
scale invariance of the return distributions (see also
~\cite{gandhi-condmat2002}).  The lack of agreement for small monthly
returns is due to short-range (1~day) correlations in the signs of the
returns that are neglected in our model, but that could easily be
incorporated~\cite{gandhi-condmat2002}. Compared to other models of
stock returns, this model shows remarkably good agreement with real
data.

We next compare these probability distributions of real and artificial
data with a ``theoretical'' prediction (solid line in Figs.~\ref{f2}).  
The model proposed leads to a 
distribution that is a Gaussian $P_{\mbox{\scriptsize t}}$ with
log-normally distributed local variance.  We can find 
$P_{\mbox{\scriptsize t}}$
by convolving Gaussians of varying widths:
\begin{equation}
P_{\mbox{\scriptsize t}}
(r)=\int\rho(\sigma)\frac{e^\frac{-r^2}{2\sigma^{2}}}{\sqrt{2\pi}\sigma}d\sigma
\end{equation}

\begin{equation}
\rho(\sigma)=\frac{1}{\sqrt{2\pi}s\sigma}e^{\frac{-1}{2}(\frac{\log\sigma-m}{s})^2}
\end{equation}
We have used the empirically found values 
$s=0.41$ and $m=-0.34$. The solid
lines in Figs.~\ref{f2} leave no doubt that this distribution is one of
the best candidates---if not the best---for describing the
distribution of price variations.  The agreement found is
exceptionally strong.

\begin{figure}
\vspace{.2in}
\centerline{\psfig{figure=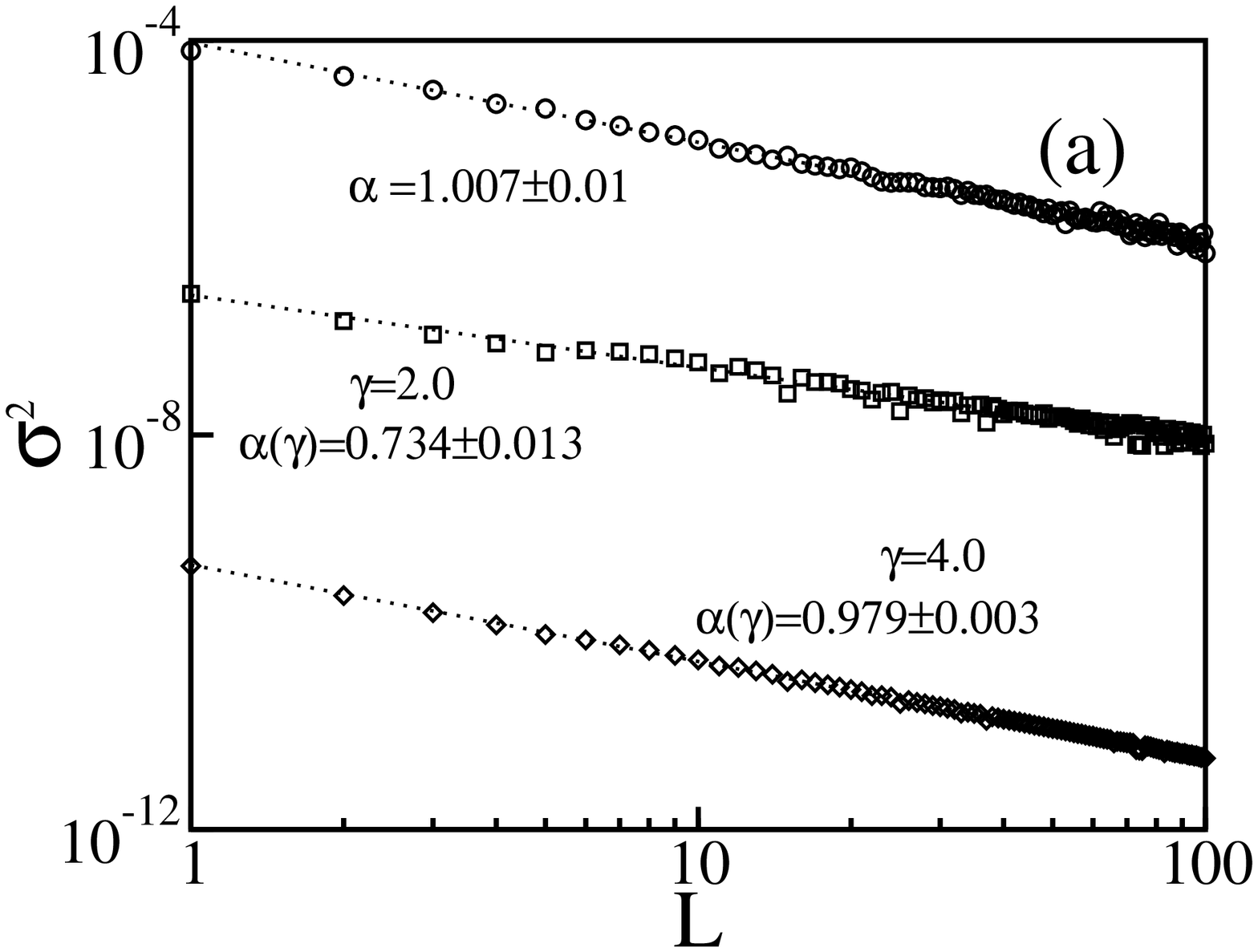,width=3.7truein}}
\centerline{\psfig{figure=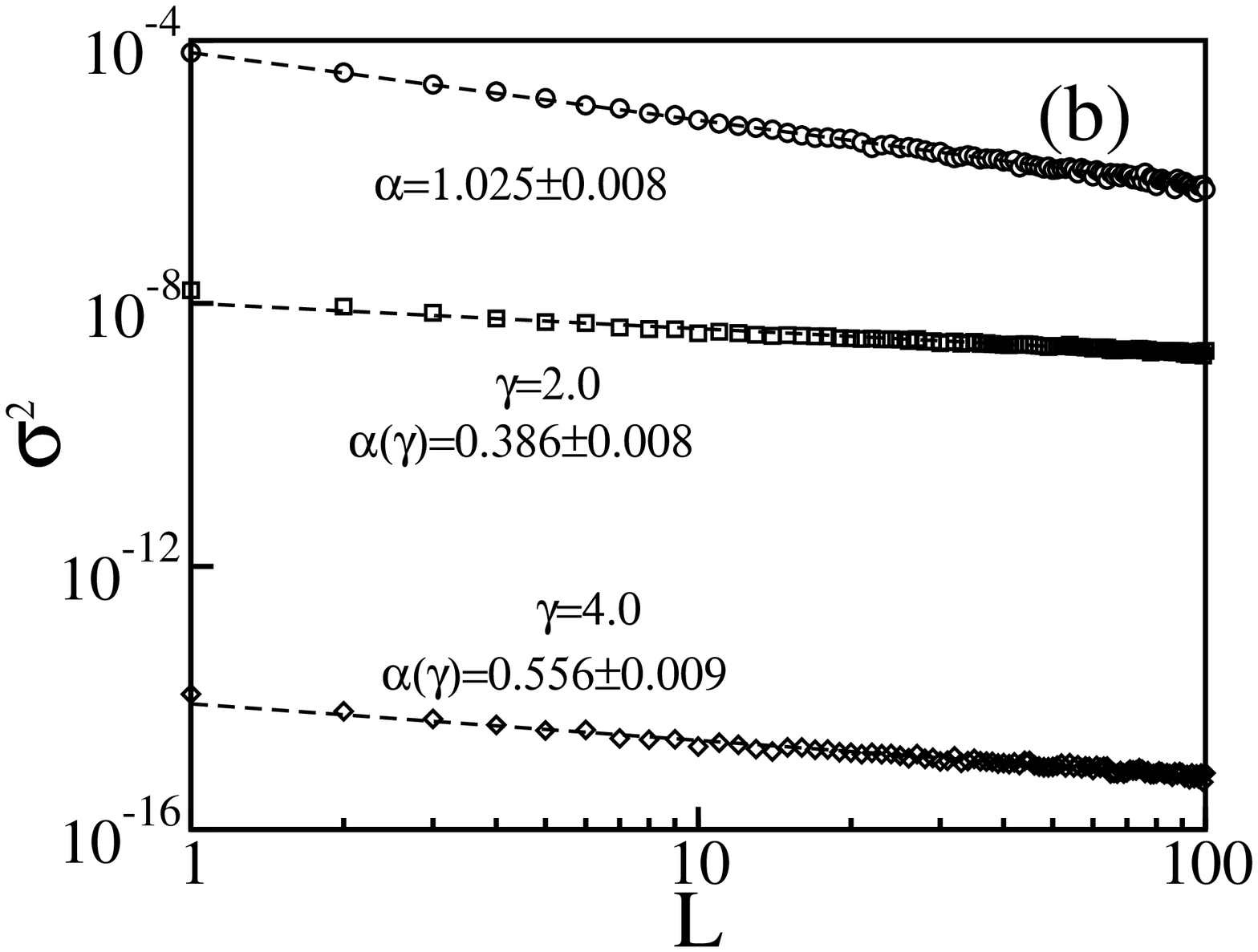,width=3.7truein}}
\bigskip
\caption{Variance $\sigma^2(\phi(L,\gamma))\sim ~ L^{-\alpha(\gamma)}$
of the generalized cumulative absolute returns as a function of $L$ on
double log scales for (a) the NYSE index and (b) the proposed
kinematic model.  Data for the absolute moments with $\gamma=2$
(square) and $\gamma=4$ (diamond) are compared with the variance
$\sigma^2(\phi(L))\sim ~ L^{-\alpha}$ of the cumulative returns
(circles).  Both data and model clearly show multifractal behavior,
since there is no unique scaling exponent.  The exponents of the best
fitting straight lines (dashed lines) are, respectively:
$\alpha(2)=0.734\pm0.013, \alpha(4)=0.979\pm0.003$ and
$\alpha=1.007\pm0.01$ for the NYSE index; $\alpha(2)=0.386\pm0.008,
\alpha(4)=0.556\pm0.009$ and $\alpha=1.025\pm0.008$ for the model.
The ability of this kinematic model to generate multifractal behavior
distinguishes it from the well-known families of models.}
\label{f3}
\end{figure}

It is known that daily returns have no auto-correlations for lags
larger than a single day, consistent with the efficient market
hypothesis. This fact can be also checked by using Detrended
Fluctuation Analysis (DFA)~\cite{pengdfa} and related
methods. Consider the cumulative returns $\phi_t(L)$, defined as the
sum of $L$ successive returns divided by $L$:
\begin{equation}
\phi_t(L)=\frac{1}{L}\sum_{i=1}^{L}r_{t+i}
\end{equation}
One can define $N/L$ non overlapping variables of this type, and
compute the associated variance $\sigma^2(\phi(L))$, where $N\approx$
9000 is the number of data points.  According to the central limit
theorem, uncorrelated (or short-range correlated) $r_{t}$ would lead
to power-law behavior: 
\begin{equation}
\sigma^2(\phi(L))\sim ~ L^{-\alpha}\;\; ,
\end{equation}
with exponent $\alpha=1$ for large $L$. The
exponent $\alpha$ both for the NYSE index and the model proposed here
is near $1$ (see Fig.~\ref{f3}), confirming that returns are
uncorrelated.

The lack of correlations does not hold true for quantities related to
the {\it absolute} returns. In order to perform the appropriate scaling
analysis, we introduce the generalized cumulative absolute returns
defined as the sum of $L$ successive absolute return powers
$|r_t|^\gamma,...,|r_{t+L-1}|^\gamma$, divided by $L$
\begin{equation}
\phi_t(L,\gamma)=\frac{1}{L}\sum_{i=1}^{L}|r_{t+i}|^\gamma
\end{equation}
where $\gamma$ is a real exponent (noting that these quantities do not
overlap.)
Using this method, if
the autocorrelation for powers of absolute returns exhibits a power-law
with exponent $\alpha(\gamma)\leq 1$ for large $L$, then we expect
\begin{equation}
\sigma^2(\phi(L,\gamma))\sim ~ L^{-\alpha(\gamma)}\;\;.
\end{equation}
(Note, however, that if the $|r_t|^\gamma$ are short-range correlated
or power-law correlated with an exponent $\alpha(\gamma)>1$, then we
would not detect anomalous scaling in the analysis of variance, because
it is not possible to detect $\alpha>1$ using the method discussed
here).

We find that both the real data and the model show similar
multifractal behavior with non-unique scaling exponents, i.e.,
$\alpha(\gamma)\neq$ constant.  Note that the function
$\alpha(\gamma)$ is not universal~\cite{lrc1,lrc2} but depends on the
particular asset considered.  Moreover, the values $\alpha(\gamma)$ we
compute do not coincide for real and artificial data, nevertheless
this is not a drawback if one takes into account the non-universality.
For completeness, we note that in order to obtain full agreement for
the multifractal exponents of any specific economic series one must
generalize the volatility as follows:
\begin{equation}
x_{t+1}= e^{a+b\omega_{t+1}} ~ (\overline{x^c_t})^d.
\label{e1}
\end{equation}
where the exponent $c$ in the volatility average would allow
for a finer control of the effects caused by extremal events. Also
the resulting returns could be written as
\begin{equation}
r_t=(\alpha+x_t) \cdot \tilde{\omega}_t\equiv \sigma_t ~
\tilde{\omega}_t.
\label{e3}
\end{equation}
with $\alpha$ representing an intrinsic constant contribution to
daily volatility.


We now comment on the motivation for using two independent Gaussian
variables, one for returns and one for volatility.  
Indeed, this choice separates the temporal evolution of volatility and
returns.  Besides giving reasonable agreement with experimental data,
there is a deeper underlying motivation for this approach.  In
ARCH-GARCH models the present-day volatility depends on the previous
day's absolute return.  Any trader knows that today's sentiment about
volatility does not depend on the previous day's variation of price
but rather on other things like previous day intraday volatility,
implicit volatility in derivative products and public (or less public)
information.  Therefore, the previous day's absolute return (which can
be small after a fearful market day with enormous variation of prices)
does not directly influence the present day volatility and so the
evolution of the latter should be kept separate, exactly as we have
done in the model proposed.

In conclusion, the proposed stochastic multifractal model of
long-range correlated absolute returns is able to reproduce features
of real financial data that are not well accounted for by existing
models.  Specifically, our model is able to overcome the inability of
ARCH-GARCH and L\'evy models to adequately explain (i)
the fat tailed distribution of the returns and its time scaling, (ii)
the long-range power law volatility correlations, or (iii) the
multifractality of the returns.  The model presented here also
represents an advance due to the exceptionally improved agreement with
real data.  We hope that this advance in modeling the kinematics of
financial time series further contributes to the emerging study of
econophysics and towards a better understanding of wider classes of
complex systems.

\bigskip
We thank Michele Pasquini for useful discussions and for a
critical reading of the manuscript. This work was supported by the
Brazilian research agencies CAPES and CNPq and by the Alagoas State
agency FAPEAL.

\newpage

\end{document}